\magnification 1200

%
%
\newdimen\FigSize       \FigSize=.9\hsize 
%
\newskip\abovefigskip   \newskip\belowfigskip
\gdef\epsfig#1;#2;{\par\vskip\abovefigskip\penalty -500
   {\everypar={}\epsfxsize=#1\nd
    \centerline{\epsfbox{#2}}}%
    \vskip\belowfigskip}%
%
\newskip\figtitleskip
\gdef\tepsfig#1;#2;#3{\par\vskip\abovefigskip\penalty -500
   {\everypar={}\epsfxsize=#1\nd
    \vbox
      {\centerline{\epsfbox{#2}}\vskip\figtitleskip
       \centerline{\figtitlefont#3}}}%
    \vskip\belowfigskip}%
%
\newcount\FigNr \global\FigNr=0
\gdef\nepsfig#1;#2;#3{\global\advance\FigNr by 1
   \tepsfig#1;#2;{Figure\space\the\FigNr.\space#3}}%
%
%
%
\gdef\ipsfig#1;#2;{
   \midinsert{\everypar={}\epsfxsize=#1\nd
              \centerline{\epsfbox{#2}}}%
   \endinsert}%
%
\gdef\tipsfig#1;#2;#3{\midinsert
   {\everypar={}\epsfxsize=#1\nd
    \vbox{\centerline{\epsfbox{#2}}%
          \vskip\figtitleskip
          \centerline{\figtitlefont#3}}}\endinsert}%
%
\gdef\nipsfig#1;#2;#3{\global\advance\FigNr by1%
  \tipsfig#1;#2;{Figure\space\the\FigNr.\space#3}}%
\newread\epsffilein    
\newif\ifepsffileok    
\newif\ifepsfbbfound   
\newif\ifepsfverbose   
\newdimen\epsfxsize    
\newdimen\epsfysize    
\newdimen\epsftsize    
\newdimen\epsfrsize    
\newdimen\epsftmp      
\newdimen\pspoints     
\pspoints=1bp          
\epsfxsize=0pt         
\epsfysize=0pt         
\def\epsfbox#1{\global\def\epsfllx{72}\global\def\epsflly{72}%
   \global\def\epsfurx{540}\global\def\epsfury{720}%
   \def\lbracket{[}\def\testit{#1}\ifx\testit\lbracket
   \let\next=\epsfgetlitbb\else\let\next=\epsfnormal\fi\next{#1}}%
\def\epsfgetlitbb#1#2 #3 #4 #5]#6{\epsfgrab #2 #3 #4 #5 .\\%
   \epsfsetgraph{#6}}%
\def\epsfnormal#1{\epsfgetbb{#1}\epsfsetgraph{#1}}%
\def\epsfgetbb#1{%
%
%
\openin\epsffilein=#1
\ifeof\epsffilein\errmessage{I couldn't open #1, will ignore it}\else
%
%
   {\epsffileoktrue \chardef\other=12
    \def\do##1{\catcode`##1=\other}\dospecials \catcode`\ =10
    \loop
       \read\epsffilein to \epsffileline
       \ifeof\epsffilein\epsffileokfalse\else
%
%
          \expandafter\epsfaux\epsffileline:. \\%
       \fi
   \ifepsffileok\repeat
   \ifepsfbbfound\else
    \ifepsfverbose\message{No bounding box comment in #1; using
defaults}\fi\fi
   }\closein\epsffilein\fi}%
%
%
\def\epsfsetgraph#1{%
   \epsfrsize=\epsfury\pspoints
   \advance\epsfrsize by-\epsflly\pspoints
   \epsftsize=\epsfurx\pspoints
   \advance\epsftsize by-\epsfllx\pspoints
%
%
   \epsfxsize\epsfsize\epsftsize\epsfrsize
   \ifnum\epsfxsize=0 \ifnum\epsfysize=0
      \epsfxsize=\epsftsize \epsfysize=\epsfrsize
%
arithmetic!
%
     \else\epsftmp=\epsftsize \divide\epsftmp\epsfrsize
       \epsfxsize=\epsfysize \multiply\epsfxsize\epsftmp
       \multiply\epsftmp\epsfrsize \advance\epsftsize-\epsftmp
       \epsftmp=\epsfysize
       \loop \advance\epsftsize\epsftsize \divide\epsftmp 2
       \ifnum\epsftmp>0
          \ifnum\epsftsize<\epsfrsize\else
             \advance\epsftsize-\epsfrsize \advance\epsfxsize\epsftmp
\fi
       \repeat
     \fi
   \else\epsftmp=\epsfrsize \divide\epsftmp\epsftsize
     \epsfysize=\epsfxsize \multiply\epsfysize\epsftmp
     \multiply\epsftmp\epsftsize \advance\epsfrsize-\epsftmp
     \epsftmp=\epsfxsize
     \loop \advance\epsfrsize\epsfrsize \divide\epsftmp 2
     \ifnum\epsftmp>0
        \ifnum\epsfrsize<\epsftsize\else
           \advance\epsfrsize-\epsftsize \advance\epsfysize\epsftmp \fi
     \repeat
   \fi
%
%
   \ifepsfverbose\message{#1: width=\the\epsfxsize,
height=\the\epsfysize}\fi
   \epsftmp=10\epsfxsize \divide\epsftmp\pspoints
   \vbox to\epsfysize{\vfil\hbox to\epsfxsize{%
      \includegraphics{#1}%
      \hfil}}%
\epsfxsize=0pt\epsfysize=0pt}%
%
%
{\catcode`\%=12
\global\let\epsfpercent=
%
%
\long\def\epsfaux#1#2:#3\\{\ifx#1\epsfpercent
   \def\testit{#2}\ifx\testit\epsfbblit
      \epsfgrab #3 . . . \\%
      \epsffileokfalse
      \global\epsfbbfoundtrue
   \fi\else\ifx#1\par\else\epsffileokfalse\fi\fi}%
%
%
\def\epsfgrab #1 #2 #3 #4 #5\\{%
   \global\def\epsfllx{#1}\ifx\epsfllx\empty
      \epsfgrab #2 #3 #4 #5 .\\\else
   \global\def\epsflly{#2}%
   \global\def\epsfurx{#3}\global\def\epsfury{#4}\fi}%
%
%
\def\epsfsize#1#2{\epsfxsize}%
%
%

\epsfverbosetrue                        
\abovefigskip=\baselineskip             
\belowfigskip=\baselineskip             
\global\let\figtitlefont\bf             
\global\figtitleskip=.5\baselineskip    
\font\tenmsb=msbm10   
\font\sevenmsb=msbm7
\font\fivemsb=msbm5
\newfam\msbfam
\textfont\msbfam=\tenmsb
\scriptfont\msbfam=\sevenmsb
\scriptscriptfont\msbfam=\fivemsb
\def\Bbb#1{\fam\msbfam\relax#1}
\let\nd\noindent 

\def\natural{{\rm I\kern-.18em N}}

\def\S{{\cal S}}
\def\R{{\Bbb R}}

\def\mylesssim{\lower.8ex\hbox{$\ \mathop{\buildrel {\textstyle <}\over 
\sim}\nolimits\ $}}
\def\mygtrsim{\lower.8ex\hbox{$\ \mathop{\buildrel {\textstyle >}\over 
\sim}\nolimits\ $}}
\def\mysmalllesssim{\lower.8ex\hbox{$\mathop{\buildrel {\scriptstyle 
<}\over {\scriptstyle\sim}}\nolimits$}}
\def\mysmallgtrsim{\lower.8ex\hbox{$\mathop{\buildrel {\scriptstyle 
 >}\over {\scriptstyle\sim}}\nolimits$}}
\def\chix{{\raise.5ex\hbox{$\chi$}}}
\def\chixa{{\chix\lower.2em\hbox{$_A$}}}

\def\real{{\rm I\kern-.2em R}}
\def\integer{{\rm Z\kern-.32em Z}}
\def\complex{\kern.1em{\raise.47ex\hbox{
            $\scriptscriptstyle |$}}\kern-.40em{\rm C}}
\def\vs#1 {\vskip#1truein}
\def\hs#1 {\hskip#1truein}

\def\Month{\ifcase\number\month \relax\or January \or February \or
  March \or April \or May \or June \or July \or August \or September
  \or October \or November \or December \else \relax\fi }
\def\date{\Month \the\day, \the\year}

  \hsize=6.5truein        
  \vsize=8.8truein      
  \pageno=1     \baselineskip=11.6pt
  \parskip=0 pt         \parindent=20pt
  \overfullrule=0pt     \lineskip=0pt   \lineskiplimit=0pt
  \hbadness=10000 \vbadness=10000 
\centerline{Wulff shape for equilibrium phases}
\vskip .1truein\centerline{by}
\vskip .1truein
\centerline{{Charles Radin
\footnote{*}{Research supported in part by NSF Grant DMS-1509088}}}
\vskip .2truein\centerline{Department of Mathematics} 
\vskip 0truein\centerline{University of  Texas} 
\vskip 0truein\centerline{Austin, TX\ \ 78712}
\vs.5
\vs1 \nd
\centerline{Abstract}
\vs.3
We use surface tension to distinguish
between phases with isotropic internal structure from phases which are
microscopically anisotropic. There are many interesting open
problems, especially in two dimensions, and in phase coexistence.
\vfill \eject
The sphere is the shape which minimizes the surface tension of a drop
of fluid. On the molecular scale a fluid -- gas or liquid -- consists of
an isotropic, disorderly configuration of particles.  The Wulff shape
traditionally plays the same minimum-surface-tension role, but for the
orderly, anisotropic particle configurations of crystals.  For an
intuitive picture consider liquid oil filling a large cylinder with a
movable piston at one end, so the oil can be subjected to a range
of pressure. The cylinder should also be temperature controlled. Now
imagine a drop of water was injected into the middle of the oil, the
water and oil being immiscible. The water would quickly attain a
spherical shape because of surface tension. (We are ignoring gravity.)
Finally, by slowly varying the temperature and/or pressure of the oil
we can change the water to ice, which is crystalline at the molecular
scale. The external shape which the ice assumes is its polyhedral Wulff
shape: the sphere of water becomes a polyhedron of ice. We emphasize
that the boundary shape is therefore an indicator of the
different internal structure between fluid water and solid ice. Indeed
the existence of facets in crystals was the inspiration for Ha\"uy to
conjecture an atomic-like structure for matter back in the eighteenth
century [H]. We will return to this point, which is the focus of this article.

Many years ago the mathematics of minimum-surface-tension surfaces was
extended within the calculus of variations from Plateau's problem, by
Jean Taylor and others [T], to include a wide range of possible
anisotropic formulas for surface tension, which one imagines are based
on a supposed underlying structure, with crystalline structures being
simple examples.  However in physics `surface tension', like
`crystal', is random at the molecular scale. We describe here an
attempt to model minimum-surface-tension shapes in this more
fundamental setting. Afterwards we contrast our approach with recent
work on Wulff shapes in the Ising model, where the objective (and
meaning) is quite different, based on different roles for randomness
in the two settings.

Consider the pressure-temperature ensemble for $N$ particles
in $\R^d$, in a container of shape $S$, with potential energy
function, for particles at $x$ and $y$, given by:

$$ v(x,y)=\cases{+\infty,&if $|x-y| < 1$;\cr |x-y|-3,&if 1 $\le
|x-y| \le 3$;\cr 0,&if $|x-y| > 3$.\cr}\eqno{1)}$$

Usually one chooses some container shape such as a unit cube, and then
the distribution is over the volume $V$ and the $2N$
position-momentum coordinates of the $N$ particles restricted to the
container scaled to have volume $V$. The distribution depends
on parameters $T,P$ as follows. 

\nd Given a function $f$ of $2m$ particle coordinates, $x_1,\ldots
x_m,p_1,\ldots p_m$, $m<N$, the integral of $f$ is:
$$\eqalign{{1\over Z_N(P,T)}\int_{(VS)^N\times \R^{dN}\times \R^+}f(x_1,\ldots x_m,p_1,\ldots p_m)&
e^{-\beta E(x_1,\ldots x_N,p_1,\ldots p_N)}e^{-\beta PV}\cr
&dx_1\ldots dx_N dp_1\ldots dp_N dV,} \eqno{2)}$$

$$E(x_1,\ldots x_N,p_1,\ldots p_N)=\sum_{j=1}^N
p_j^2/2+\sum_{j,k=1}^Nv(x_j,x_k), \eqno{3)}$$

\nd $\beta=1/(k_BT)$, \ and $Z_N(P,T)$ is the normalization constant 
(partition function):

$$Z_N(P,T)=\int_{(VS)^N\times \R^{dN}\times \R^+}
e^{-\beta E(x_1,\ldots x_N,p_1,\ldots p_N)}e^{-\beta PV}dx_1\ldots
dx_Ndp_1\ldots dp_N dV. \eqno{4)}$$

Although it has not been proven, it is expected that in the limit of
large $N$ this standard formalism will exhibit the familiar
thermodynamic phases as in Figure 1, at least for $d=3$. We elaborate on that in
the context of a new twist. 
\vs.1
\hs1 \hbox{\epsfxsize=3.5in \epsfbox{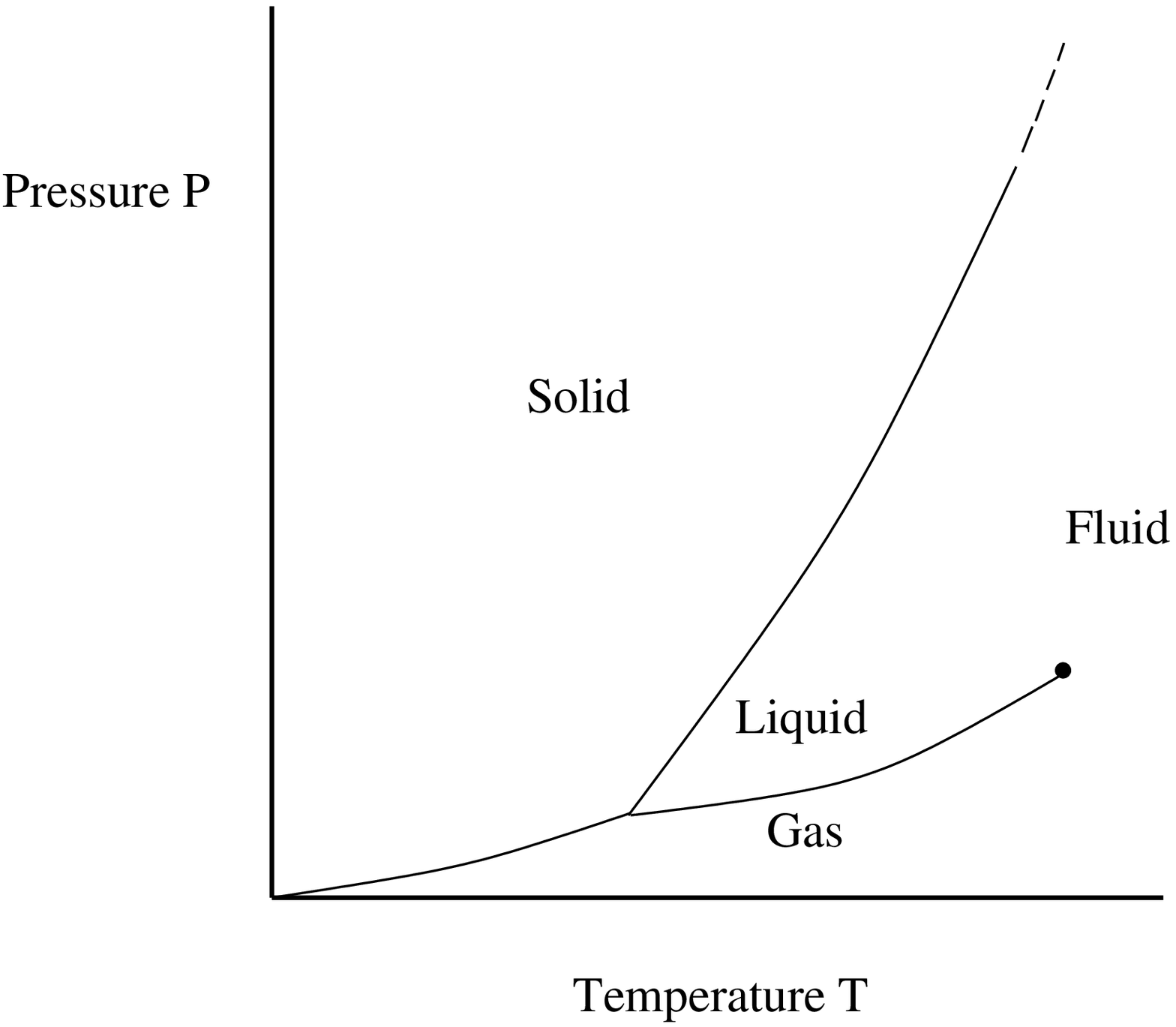}}

\centerline{Figure 1. The phase diagram of a simple material.}

\vs.2 

As a new feature we propose to treat the (star-shaped, fixed
volume) container shape $S$ as a variable.  For fixed $S$ it is
expected that for `most' pairs $(T,P)$, i.e.\ except for phase
coexistence on the phase transition curves in Figure 1, there is a
unique Euclidean invariant weak-limit Gibbs probability measure as
$N\to \infty$, which is {\it independent of} $S$. This is why $S$ is
not usually considered an interesting variable. However the phenomenon
we wish to model is a surface effect, so it is necessary to 
analyze finite systems before taking any infinite system limit. (This
approach is related to the one in [AR].)

For each $S$ the pressure-temperature ensemble defined by 2) is a
probability distribution over particle configurations. If we think of
$S$ as a free surface, the minimization of surface tension can be
implemented by minimizing the average total energy of the
configurations: particles away from the boundary obtain a
certain negative potential energy from their surrounding particles,
while those near the surface have a higher potential energy -- losing
the negative potential energy of missing neighbors -- so minimizing the
energy should tend to minimize surface area. This is the underlying nature of
`surface tension'. Now this should lead to $S$ being a sphere if most
configurations are isotropic, as in a fluid, but not if they are
ordered, as in a crystal. This is the idea, but how does one implement
a free surface? 

If we model the surface as an interface between molecular water and
molecular oil we would have to show that the interface becomes sharp
under some scaling. Instead we will assume the interface is sharp but
flexible, responding to the random configuration of water
molecules. It will still be difficult to obtain a well-defined optimal
shape, as we shall see.

Consider the mean value $\langle E\rangle_{T}^P(S,N)$ of the total
energy $E$ for the system with $N$ particles, where we emphasize its
dependence on $S$ and $N$. Let us define a topological space for
$S$. Since we are only interested in star-shaped $S$, we model $S$ by
giving the radial coordinate $\rho$ as a positive continuous function
of `angle' $\alpha$, i.e.\ a point on the compact unit sphere
$S^d$. More specifically, starting with the set of continuous
functions on $S^d$ with values $\rho(\alpha)\ge 1$, we then define the
subset associated with volume $10$ (computed in spherical coordinates
from $\rho$), and finally define our space of shapes as the quotient
$\S$ of that space by Euclidean motions, in the quotient topology. For
each point in $\S$ we have defined in 2), using similarities $VS$ of
$S\in \S$, the family of pressure-temperature probability
distributions, which are continuous in $S$ in the weak topology. In
particular $\langle E\rangle_{T}^P(S,N)$ is continuous in $S$. At this
point we do not see an easy way to prove the existence of the minima we want,
to eventually define a Wulff shape as a function of
$(T,P)$. Technically we now do this by replacing $\S$ by some
convenient compact subset, containing at least the sphere and some
appropriate polyhedra. This is unsatisfying, but does not impede our main
goal, as we shall see. Choosing such a
subset we find that $\langle E\rangle_{T}^P(S,N)$ achieves a minimum
at one or more $S=S_W(P,T,N)$. We denote by $S_W(P,T)$ any
accumulation point, as $N\to \infty$, of these $S_W(P,T,N)$, and call
it a Wulff shape for $(T,P)$.

For $d=3$, large fixed $N$, any fixed $T>0$ and $S$, and all large enough
$P$ (high density) i.e.\ above the solid/fluid transition in Figure 1,
the pressure-temperature distribution is expected to concentrate on
`crystal-like' particle configurations, while for all small enough $P$
(low density) i.e.\ below the transition, it is expected to
concentrate on highly disordered configurations, with a sharp
transition between the two regimes as $N\to \infty$,
defining the transition curve. Therefore fixing $T$ and
large $N$, we expect for all large enough $P$ that $S_{W}(P,T,N) $
would be approximately a polyhedral shape (cuboctahedron) which minimizes the
boundary for
densest packing of spheres [BoRa] (and $S_W(P,T)$ to be exactly this), while
for all small enough $P$, we expect $S_W(P,T,N)$ to be approximately
the sphere (and $S_W(P,T)$ to be exactly this), since it minimizes the
energy due to particles near the surface.  In both cases this is a
surface effect.

Needless to say these conjectures are far from proven; after all, they
imply a solid/fluid phase transition, something that has so far
resisted proof in any reasonable model; see however [BLRW]. In fact
one reason for taking this path is the hope that perhaps this
could be a way to prove such a transition [AR]. Indeed, if one could
simply prove that for some $(P,T)$ the sphere gives a lower value to
surface energy than some other shape, say a polyhedron, while for
other $(P,T)$ the polyhedron gives a lower value than the sphere, this
would be of great interest, and we emphasize that this does not
require proving the existance of optimal shapes.

We also note that in the above two  paragraphs we assumed $d=3$. The reason
is that for $d=2$ it is known that you cannot have a crystal in the
full sense of long-range positional order [Ri], the densest phase
presumably only having orientational long-range order [BK]. It is
therefore an interesting open problem to determine what the Wulff
shape would be for high pressure in our model in $d=2$. (The regular
hexagon is a natural candidate, and was proven for $T=P=0$ in
[Ra].)

At this point one might naturally ask what happens to the Wulff shapes
as one varies $(T,P)$ through the phase transition curve separating
the solid and fluid phases. In terms of the oil/water discussion
above, for $(T,P)$ actually on the curve separating the fluid and
solid phases one has a mixture of liquid water and a chunk of ice
coexisting together in the oil. (To fix the relative volume of each
phase one needs to use a different ensemble than the one used here.)
One could perhaps analyze the shape of the ice in contact with
(perhaps surrounded by) water, but we emphasize that this is
inherently different from the above analysis within the pure fluid and
solid phases, and is precisely the problem we avoided by not modeling
the oil at a molecular level. In particular one would have to address the
question of whether it is appropriate to consider the interface
between water and ice as well-defined, since we expect molecules to be
constantly moving from one phase-region to the other. We avoided this by
assuming (complete) immiscibility of the water and oil. This
phenomenon has
been the focus of a sequence of beautiful works in probability, as
part of an emerging subfield called `limit shapes': not however for
the interface between ice and water, but for that between water and
steam, in both of which the internal structure is isotropic. (For a
useful introduction to limit shapes, including a review of work on
Wulff shapes in the Ising model, see the Colloquium lectures of
Okounkov [O].) Assuming the reader is somewhat familiar with the Ising
model we give a very brief description of this work, with an emphasis
on contrasting it with our modeling with Wulff shapes above.

The ferromagnetic Ising model on the square lattice can be presented
(in the so-called grand canonical ensemble) in terms of a two
parameter family of probability distributions, the parameters being
temperature $T>0$ and a constant external magnetic field $B$. It
exhibits one phase, with a transition across a segment of the line
$B=0$ as indicated in Figure 2.
\vs.1
\hs1.5 \hbox{\epsfxsize=2.5in \epsfbox{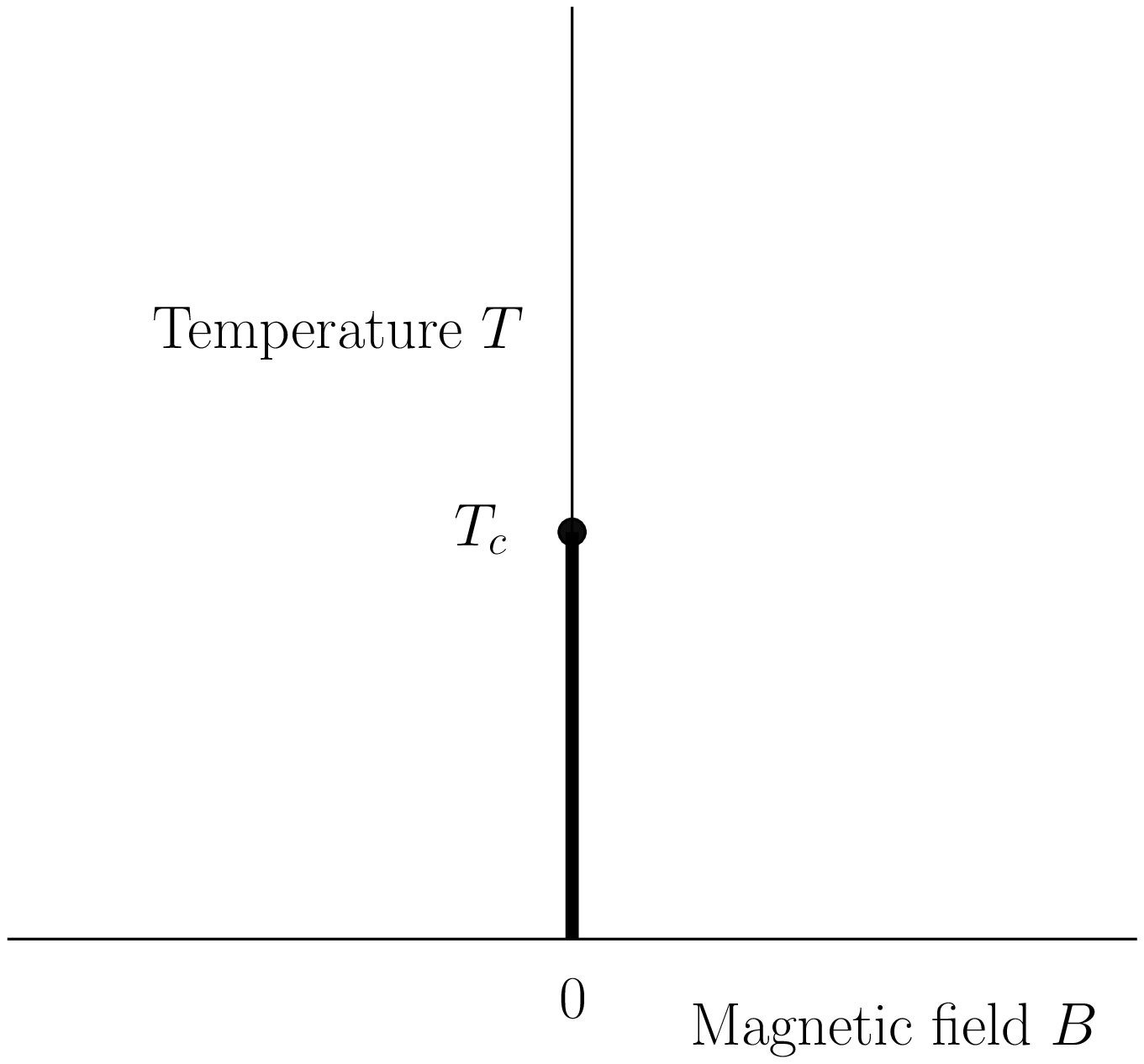}}
\vs.1
\centerline{Figure 2. The phase diagram of the Ising model.}
\vs.2

As we have done for the particle model with interaction 1), the Ising
model is usually presented first for a finite system, and then one
takes an infinite system limit. If one introduces a fixed boundary for
the finite system model using a star-shaped curve $S$ as done above
for models based on 1), one sees that for $B\ne 0$ the Wulff
shape based on minimizing energy is a square. The case $B=0$ is more
complicated and includes the temperature interval $[0,T_c)$, below the
critical point, where the limiting probability measure is not unique
and represents coexistence of states rich in spin up with states rich
in spin down. When $B<0$ one only gets the spin up type and when $B>0$
one only gets the spin down type, and as noted above each has a square
Wulff shape. 
One might imagine this perseveres in coexistence where
both types of states are present, but careful analysis
shows that the situation is more complicated [O].

Our reason for bringing up this work on the Ising model is to avoid
confusion between the Ising work on the shape of interfaces, for
thermodynamic parameters representing coexistence (or a `mixed
phase'), from our modeling of minimum surface energy in a pure `solid'
phase with anisotropic internal structure; the analyses are focused
on different problems but both use the term Wulff shape. If we view
the Ising model as a lattice gas following Lee-Yang [LY], the work on
Ising Wulff shape could naturally be interpreted as an analysis of the
{interface} between water and steam (both having the same isotropic Wulff
shape), in particular the scaling appropriate to making the interface
sharp, while ours does not seriously attempt to determine well-defined
optimal surfaces, and instead tries to analyze the
difference between the internal structure of water and ice, each in
its isolated state. Our motivation is that the transition between
water and steam is generally considered well understood (away from the
critical point!) as a
transition between high density and low density states of a single
fluid phase, while the transition between ice and water is not well
understood. Following Landau the solid/fluid transition is usually
associated with a change of symmetry in structure at the molecular level, and
this is used in attempts to understand the level of continuity of the 
transition (see [RRS]). This paper is an attempt to use surface
tension to distinguish such phases. 
\vs.1 \nd
In conclusion the main question we pose, for $d=2,3$, is:
\vs.05
\item{} For the model of particles interacting through 1) is surface 
tension lower (i.e.\ is the average energy lower)
for a polyhedron than the sphere for fixed temperature and all
sufficiently high pressure?

\vs.2
\centerline{Bibliography}
\vs.2
\item{[AR]} D.\ Aristoff and C.\ Radin, Rigidity in solids, 
J.\ Stat.\ Phys.\ 144 1247-1255 (2011).

\item{[BK]} E.\ P.\ Bernard and W.\ Krauth,
Two-step melting in two dimensions: first-order liquid-hexatic transition
Phys.\ Rev.\ Lett.\ 107 155704 (2011).

\item{[BoRa]} B.\ Borden and C.\ Radin, The crystal structure of the noble gases,
J.\ Chem.\ Phys.\ 75 2012-2013 (1981).

\item{[BLRW]} L.\ Bowen, R.\ Lyons, C.\ Radin, and P.\ Winkler, 
Fluid-Solid Transition in a Hard-Core System, 
Phys.\ Rev.\ Lett.\ 96 025701 (2006).

\item{[H]} R.\ J.\ Ha\"uy, Essai d'une Th\'eorie sur la Structure des Crystaux,
Paris: Chez Gogu\'e \& N\'ee de la Rochelle, (1784).
\vfill \eject 
\item{[LY]} T.\ D.\ Lee and C.\ N.\ Yang, statistical theory of equations 
of state and phase transitions. II. Lattice gas and Ising model,
Phys.\ Rev.\ 87 410-419 (1952).

\item{[O]} A.\ Okounkov, Limit shapes, real and imagined,
Bull.\ Amer.\ Math.\ Soc.\ 53 187-216 (2016).

\item{[Ra]} C.\ Radin, The ground state for soft disks, J. Stat. Phys.\ 26
365-373 (1981). 

\item{[Ri]} T.\ Richthammer,
Translation-invariance of two-dimensional gibbsian point processes,
Commun.\ Math.\ Phys.\ 274 81-122 (2007).

\item{[RRS]} C.\ Radin, K.\ Ren and L.\ Sadun, A symmetry breaking transition 
in the edge/triangle network model, arXiv:1604.07929v1.

\item{[T]} J.\ E.\ Taylor, Crystalline variational methods,
Proc.\ Nat.\ Acad.\ Sci.\ 99 15277-15280 (2002)

\end